\documentclass[preprintnumbers,nofootinbib,secnumarabic]{revtex4}
\usepackage{amsmath,graphicx,amssymb,multirow,bbm}

\usepackage{epsfig, subfigure}
\usepackage{slashed}
\usepackage[english]{babel}

\usepackage{color}

\def\ga{\mathrel{\raise.3ex\hbox{$>$\kern-.75em\lower1ex\hbox{$\sim$}}}}
\def\la{\mathrel{\raise.3ex\hbox{$<$\kern-.75em\lower1ex\hbox{$\sim$}}}}

\def\lsim{\mathrel{\rlap{\lower4pt\hbox{\hskip1pt$\sim$}}
    \raise1pt\hbox{$<$}}}                
\def\gsim{\mathrel{\rlap{\lower4pt\hbox{\hskip1pt$\sim$}}
    \raise1pt\hbox{$>$}}}                

\makeatletter
\renewcommand\appendix{\par
  \setcounter{section}{0}%
  \setcounter{subsection}{0}%
  \renewcommand\thesection{\@Alph\c@section}
  \setcounter{figure}{0}%
  \setcounter{table}{0}%
  \renewcommand\thefigure{\thesection.\@arabic\c@figure}
  \renewcommand\thetable{\thesection.\@arabic\c@table}
  }
\makeatother

\begin{document}

\title{Limits on strong FCNC top couplings at the LHC}

\author{Renato Guedes}
    \email[E-mail: ]{renato@cii.fc.ul.pt}
\affiliation{Centro de F\'{\i}sica Te\'{o}rica e Computacional,
    Faculdade de Ci\^{e}ncias,
    Universidade de Lisboa,
    Av.\ Prof.\ Gama Pinto 2,
    1649-003 Lisboa, Portugal}
\author{Rui Santos}
    \email[E-mail: ]{rsantos@cii.fc.ul.pt}
\affiliation{ISEL - Instituto Superior de Engenharia de Lisboa,
	1959-007 Lisboa, Portugal}
\affiliation{Centro de F\'{\i}sica Te\'{o}rica e Computacional,
    Faculdade de Ci\^{e}ncias,
    Universidade de Lisboa,
    Av.\ Prof.\ Gama Pinto 2,
    1649-003 Lisboa, Portugal}
\author{Miguel Won}
    \email[E-mail: ]{miguel.won@coimbra.lip.pt}
\affiliation{LIP/Departamento de F\'{\i}sica,  \\
Universidade de Coimbra, 3004-516 Coimbra, Portugal}

\date{\today}

\begin{abstract}
The best limit on the strong FCNC anomalous couplings was obtained using the direct top
production process at the Large Hadron Collider by the ATLAS collaboration.
We perform a similar analysis but using a next-to-leading order generator, MEtop.
We then show how the limits could be improved if the FCNC single top process $pp \to tj$
would be included as signal. Finally we discuss a slightly modified analysis with an extra
hard jet in the final state.
\end{abstract}

\maketitle

\section{Introduction}
The Large Hadron Collider (LHC) has now concluded the 8 TeV run gathering a total of approximately 20 fb$^{-1}$ 
of integrated luminosity. The large
number of top-quarks produced allow for an unprecedented precision in the study of top quark observables.
It is well known that Flavour Changing Neutral Currents (FCNC) involving a top quark are negligibly small in the
Standard Model (SM). They can however be larger by up to eight orders of magnitude in some extensions
of the SM~\cite{AguilarSaavedra:2004wm}. Therefore, the search for signals of FCNC processes involving the top-quark
is an excellent means of testing the validity of the SM while simultaneously probing some of its extensions. Direct top production is 
the most sensitive process to probe the FCNC vertex involving a top-quark, a light-quark and a gluon. At the parton level, direct
top is simply given by $g q \, (\bar q)  \to t \, (\bar t)$ where $q=u,c$ and $g$ is a gluon, $u, \, c$ and $t$ are the up, charm and
top quarks respectively.

No evidence for top-quarks originating from the direct top process was found in the searches 
performed by the experiments at the Tevatron~\cite{Aaltonen:2008qr} and at the LHC
where the best limits on the anomalous strong FCNC couplings were obtained by the 
ATLAS collaboration~\cite{Aad:2012gd}. 
In the ATLAS analysis~\cite{Aad:2012gd}, the direct top signal events were generated with
the PROTOS~\cite{Saavedra:2013} Monte Carlo generator. The upper limit on the production 
cross-section was then converted into limits on the anomalous coupling constants.
In order to account for the next-to-leading-order (NLO) corrections,  a K-factor was used according to the NLO calculation
of the FCNC direct top process~\cite{Liu:2005dp, Gao:2011fx}. The bound can also be written as
a 95 \% C.L. limit on the top FCNC branching ratios $Br(t \rightarrow ug) < 5.7\times 10^{-5} $
and $Br(t \rightarrow cg) <2.7\times 10^{-4} $~\cite{Aad:2012gd} with the assumption that only one of the 
FCNC vertices $g u \, \bar t$ or $g c \, \bar t$ is turned on at a time. In this work we perform a similar analysis
using the NLO generator MEtop~\cite{Coimbra:2012ys}, a tool made recently available dedicated to top quark FCNC production.
MEtop has a complete set of dimension six operators for the study of FCNC top-quark interactions~\cite{alot}.  

The purpose of this paper is threefold. First we want to compare the previous analysis where the events were generated 
at leading-order (LO) and normalized with a K-factor with one where the direct top events are generated at NLO using MEtop. 
We will therefore redo the analysis performed by the ATLAS collaboration in~\cite{Aad:2012gd} 
with the NLO events generated by MEtop.
%
%
Second, because MEtop also includes the LO hard FCNC process $pp \to t j$, where $j$ is a light jet, we will account
for the contribution of the hard process to the analysis already performed. Our goal is to check weather a sizeable
improvement in the limit is obtained just by adding the events from the FCNC single top process to the direct top events. 
Finally, we will perform an analysis where again we follow~\cite{Aad:2012gd}  but allow for one extra hard jet in the signal.
It is clear that the final state will then be very similar to the SM single top one. Hence, our objective is to check if
the major increase in the background can be compensated by the increment in the number of signal events.

\section{Data Sample}
\label{sec:signal}
Three different sets of signal events were generated with METop~\cite{Coimbra:2012ys}, 
\begin{itemize}
\item FCNC direct top  @LO:   $Dtop^{LO}$,
\item FCNC direct top  @NLO: $Dtop^{NLO}$,
\item FCNC direct top  @NLO plus FCNC single top @LO:  $Dtop^{NLO}+(t+j)^{LO}.$
\end{itemize}
The last set $(Dtop^{NLO} + (t+j)^{LO})$ is a weighed combination
of direct top production at NLO with FCNC single top production at LO. As discussed in~\cite{Coimbra:2012ys},
only one FCNC operator for each light quark ($u$ and $c$) contributes to the direct top process. Therefore, each set
is composed of two samples - one where only the $ugt$ coupling is turned on and the other where only the $cgt$ coupling
is turned on. The generation of the FCNC single top quark events in $Dtop^{NLO}+(t+j)^{LO}$ followed
the same rule. 
All events were generated assuming a SM top quark decay, \textit{i.e.}, $BR(t \to  W^+b) \approx 100$ \% and only
the leptonic decay of the $W$ was considered. Additionally, the full $\tau$ leptonic decay was taken into
consideration in both signal and background. We have used the Parton Density Function (PDF) set CTEQ6~\cite{Pumplin:2002vw} 
for all leading order (LO) processes and CTEQ6.6~\cite{Pumplin:2002vw} for the NLO cross sections.

As previously stated the SM FCNC cross section is negligible due to its very low cross section.
The most significant backgrounds are single top production, $t\bar t$ production, $W/Z$ plus jets (both light and heavy jets), 
diboson production and the multijet background.
The single top background (t-channel, s-channel and $Wt$ associated production) together with $t\bar t$ were
generated with POWHEG~\cite{Alioli:2010xd} at NLO and the CTEQ6.6 NLO PDFs were used. 
For $W$ plus light jets, $Wc$ plus light jets, $Wb\bar b$, $Wc\bar c$ (plus light jets) and $Z$ plus light jets we have 
used AlpGEN~\cite{Mangano:2002ea} with the CTEQ6 LO PDFs. In all events generated with AlpGEN the jets have a 
transverse momentum above 20 GeV and $\Delta R_{jj}>0.7$. Further, in the $W$ plus jets case, the jets have $|\eta_j|<4.9$ 
and for $Z$ plus jets $|\eta_j|<2.5$. For both the $W$ and the $Z$ plus jets events, the number of jets was varied from 0 to 3. 
To remove overlaps between $n$ and $n+1$ partons the MLM matching scheme~\cite{mlm} was used. 
The cross sections were then normalized at NLO using MCFM~\cite{Campbell:1999ah, site}. 

The events were then submitted to a parton shower simulator performed with PYTHIA 6~\cite{Sjostrand:2006za}  in order to include 
initial (ISR) and final (FSR) state radiation, as well as multiple interactions. The Perugia tune~\cite{Skands:2009zm} 
was used to handle the underlying events in POWEG while the ATLAS MC09 tune~\cite{ATLASCollaboration:2010002} 
was used for events generated with AlpGen.  Finally, both signal and background detector simulation was 
performed with Delphes~\cite{Ovyn:2009tx}, which is a framework for the fast simulation of a generic detector in a 
collider experiment. For the detector and trigger configurations, we resorted to the ATLAS default definitions in Delphes.
However, in order to reproduce the ATLAS analysis as faithful as possible we have redefined the sum of the 
$E_T$ in a cell to be calculated within a cone of $\Delta R=0.3$ for the lepton, and $\Delta R=0.4$ for the jet. 
Additionally, the b-tagging  efficiency was set to be 57\%, and the b-mistagging to 0.2\% for light-quark jets and 10\% for c-quark jets. 
These values were chosen to match the ATLAS analysis~\cite{Aad:2012gd}.   
Finally, we have not considered the diboson and multijets background which in the ATLAS analysis~\cite{Aad:2012gd}
amounts to 9 \% of the total background (the largest contribution comes from multijets with about 6.7 \%).

\section{Event selection}

As previously discussed we have performed two different analysis. The first one is similar to one 
presented in~\cite{Aad:2012gd} by  the ATLAS collaboration.
It should be noted however that besides the usual cut-based analysis, ATLAS uses a multivariate analysis technique (neural-network) 
to separate signal from background. As we will not be using this multivariate technique, our results cannot be compared 
with theirs. This is not an issue because our aim is not to compare our analysis with the experimental one but rather 
to study its performance for different sets of events generated with MEtop. The ATLAS analysis will be used as our standard analysis
because it provides the best current limits on the $ugt$ and $cgt$ strong FCNC couplings. It will also serve as a means to 
control our background. In the present work the limits on the FCNC couplings were obtained using the ATLAS cut-based 
part of the analysis plus an additional cut on the top invariant mass. 
From now on, we shall call this analysis "ATLAS" but it should be clear that this is not the ATLAS analysis performed in~\cite{Aad:2012gd}.
A detailed description of what we call the "ATLAS" analysis will be presented below.
Still in the framework of this first analysis we will consider a new set of signal events, $Dtop^{NLO}+(t+j)^{LO}$, that is,
we will add the FCNC single top to the NLO direct top. The ATLAS final state consists of one b-quark jet, one lepton (electron or muon)
 and missing energy. In the analysis, we ask for exactly zero non-b jets. However, a jet can only be identified with $p_T > 25 GeV$ 
 and $|\eta_j|<2.5$. This means that some of the events from FCNC single top will still pass the selection if the non b-jet is soft.
Hence, we will study how the inclusion of the FCNC single top events will affect the bound on the couplings.

The second analysis will be performed considering a different final state topology
with an extra hard non-b jet. In table~\ref{tab1.1} we present the total cross section
for each of the three set of events, where the FCNC coupling constants were set to either zero or 
$\kappa_{qgt}/\Lambda=0.01$ TeV$^{-1}$. The $t+j$ sample and the corresponding total cross section
is produced with a 10 GeV cut on the jet $p_T$.
For completeness we present the strong FCNC operator which we write as
\begin{equation}
 i \kappa_{ugt} \, \frac{g_s}{\Lambda} \bar{u} \lambda^a \sigma^{\mu\nu} (f_u+h_u \, \gamma_5) t  G_{\mu\nu}^a \quad + (u \leftrightarrow c) + h.c. 
\end{equation}
where $\kappa_{ugt}$ is real, $g_s$ is the strong coupling and $f_u$ and $h_u$ are complex numbers with $|f_u|^2 + |h_u|^2 = 1$ (see~\cite{Coimbra:2012ys} for a detailed
discussion relating the forms of the strong FCNC operators).
For the chosen value of the coupling constant, the FCNC single top cross section gives an additional contribution of 14\% and 27\% to the full 
NLO direct top cross section, for the $ugt$ and $cgt$ operators respectively. These extra events are kinematically similar to 
the SM single top ones and are therefore expected to be mainly located in regions discarded by the ATLAS analysis. 
Nevertheless, it is important to understand if an analysis that considers an extra hard jet can lead to an improvement  on the 
FCNC couplings limit. We will refer to this analysis as ATLAS(m). 

\begin{table}[h!]
\begin{center}
\begin{tabular}{l|c|c|c}
\hline
   & $\sigma(Dtop^{LO})$ (pb) & $\sigma(Dtop^{NLO})$ (pb) & $\sigma(Dtop^{NLO}+(t+j)^{LO})$ (pb)\\
\hline
$ugt$	&	2.245	&	2.972	& 3.374\\
$cgt$	&	0.355	&	0.567	&	0.720\\
\hline
\end{tabular}
\end{center}
\caption{$Dtop^{LO},$ $Dtop^{NLO}$ and $Dtop^{NLO} + (t+j)^{LO}$ total cross sections for $\sqrt{s} = 8$ TeV and $\kappa_{qgt}/\Lambda=0.01$ TeV$^{-1}$. \label{tab1.1}}
\end{table}
 
We have used the ATLAS default trigger card on the Delphes detector with an isolated electron with 
$p_T>25\:GeV$ or an isolated muon with $p_T>20\:GeV.$ In the analysis we demanded at
least one electron or one muon with $p_T>25\:GeV$. Exactly one reconstructed jet with $p_T>25\:GeV$
is required. This jet has to be identified as a b-quark jet (b-tagged). We excluded events with missing 
transverse energy $\slashed{E}_T<25\:GeV.$ In order to further reduce  the multijet background - 
most of it with low $\slashed{E}_T$ and low values of the reconstructed 
W-boson transverse mass $M^W_T=\sqrt{2 p_T^l \slashed{E}_T-2 (p_x^l \slashed{E}_x+p_y^l \slashed{E}_y)}$ - 
we have required $M^W_T+\slashed{E}_T>60\:GeV.$ Finally, the top-quark invariant mass is set to be above 140 GeV. 
This last cut was not implemented by ATLAS in their cut-based part of the analysis~\cite{Aad:2012gd} but it is included 
in the multivariate part.

\begin{figure}[h!]
  \begin{center}
    \includegraphics[scale=0.43]{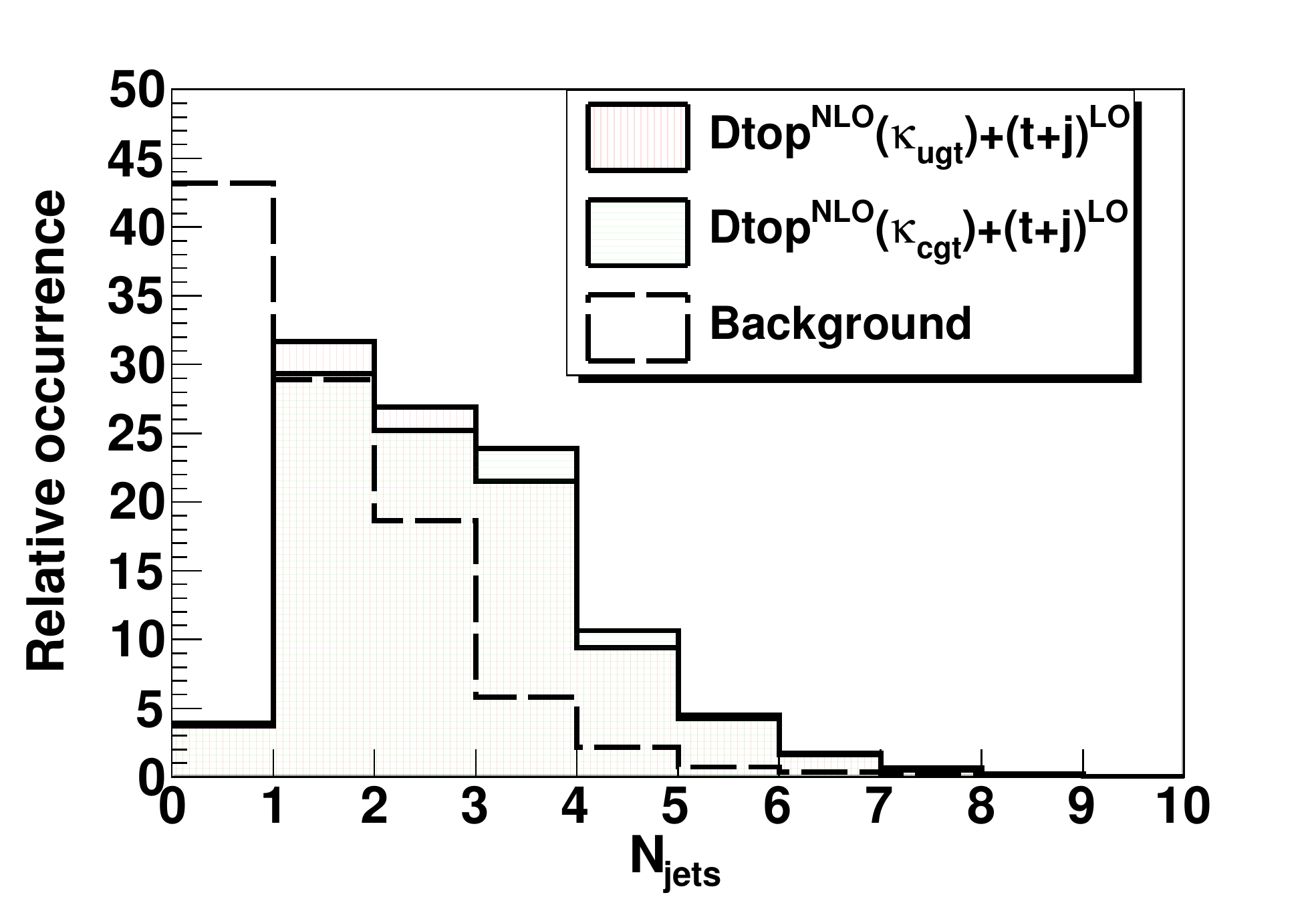}
    \includegraphics[scale=0.43]{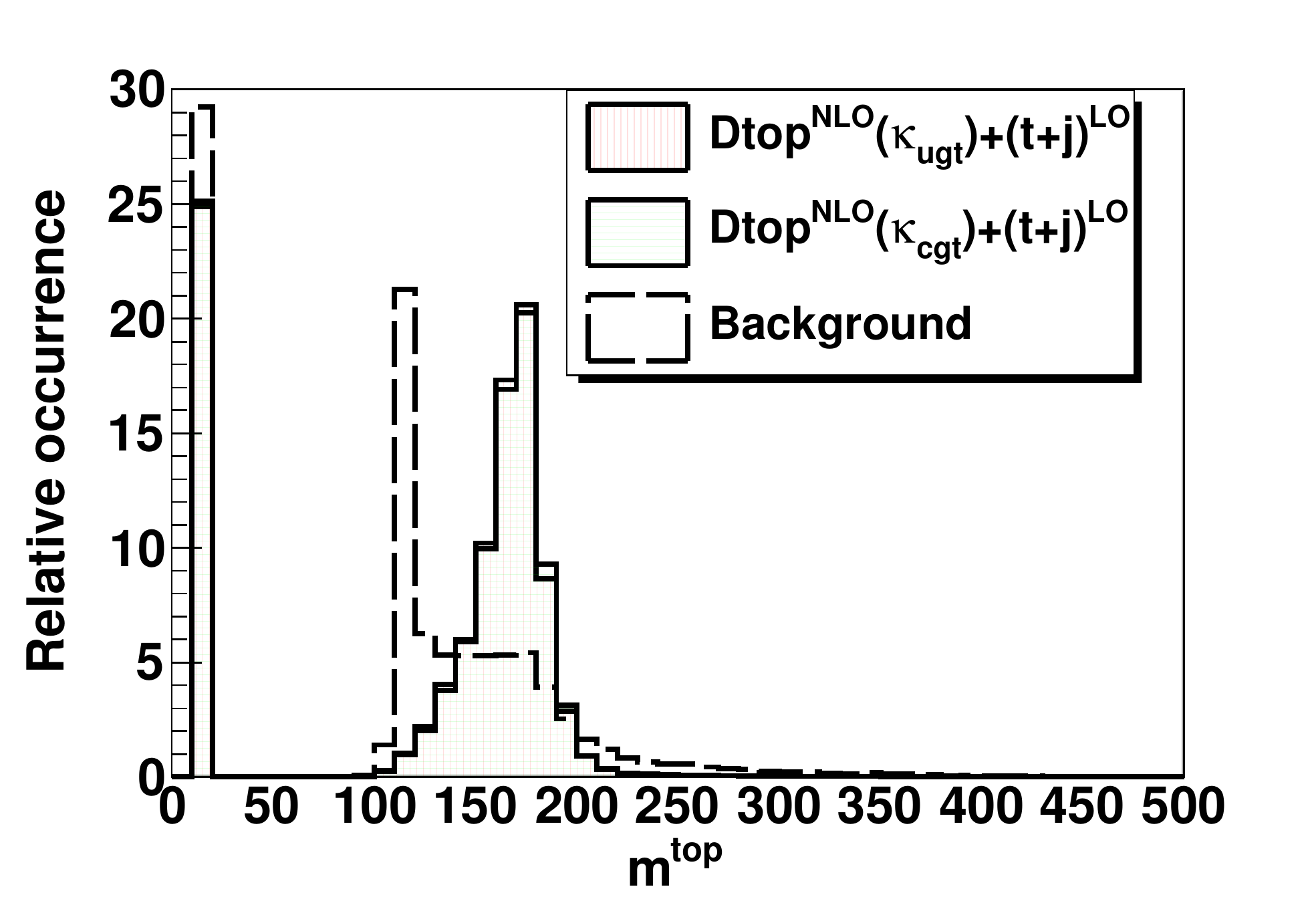}
    \caption{In the left panel we present the jet multiplicity for  jets with $p_T > 25 GeV.$ On the right we show the top-quark invariant mass.}
    \label{fig_dist}
  \end{center}
\end{figure}

In the ATLAS(m) analysis we have changed the requirements regarding jets: we have asked for one 
or two reconstructed jets with $p_T>25\:GeV$, where one jet must be a b-jet and the second is forced to be
a non-b jet. 
In the left panel of fig.~\ref{fig_dist} we show the jet multiplicity for jets with $p_T>25\:GeV$. In the
right panel we show  the top quark invariant mass before the respective  cut is implemented which allow us
to understand the effect of this additional cut in the analysis.

\begin{table}
\begin{center}
\begin{tabular}{l|cc|cc}
\hline
   & ATLAS &  &  ATLAS(m) &\\
\hline
Process	& Events $(1 fb^{-1})$ & Efficiency (\%) &  Events $(1 fb^{-1})$ & Efficiency (\%)\\
\hline
Single top												    &	330.8	    & 0.286  	&	1198.5	& 1.035	\\
$t\bar t$													&	111.0	    & 0.052	    &	773.1	& 0.365	\\
$W$ + light jets										    &	2900.1	    & 0.026  	&	4300.3	& 0.039	\\
$Wc$ + light jets										    &	1796.2	    & 0.317	    &	2384.4	& 0.421	\\
$Wb\bar b/Wc\bar c$ + light jets	            &	45.9		    & 0.591	    &	128.7	& 1.656	\\
$Z$ + jets													    &	66.4	        & 0.002	    &	116.2	& 0.004	\\
\hline
Total background	&	5250.4	&  	&	8901.2 &	\\
\hline
\end{tabular}
\end{center}
\caption{Number of events and efficiency for the background processes in the ATLAS and in the ATLAS(m) analyses. \label{tab1}}
\end{table}

In table~\ref{tab1} we list all backgrounds considered in the analysis as well as the event yield and the efficiency 
for a luminosity of $1\:fb^{-1}$. These are the final events gathered after all cuts. As expected there is a 
significant increase in the single top and $t \bar{t}$ backgrounds because there is an extra non-b jet in the
ATLAS(m) analysis. However, the overall increase in the total background in ATLAS(m) relative to ATLAS
is not as large because the major contributions to the total background comes from $W$+jets where the increment is not
so dramatic.


\begin{table}[h!]
\begin{center}
\begin{tabular}{l|c|c}
\hline
   & ATLAS &  ATLAS(m) \\
\hline
Process   & Efficiency (\%) &  Efficiency (\%)\\
\hline
$Dtop^{LO}(\kappa_{ugt})$									& 2.509		&	--\\
$Dtop^{LO}(\kappa_{cgt})$									& 3.428		&	--\\
$Dtop^{NLO}(\kappa_{ugt})$								& 2.591		&	--\\
$Dtop^{NLO}(\kappa_{cgt})$								& 3.581		&	--\\
$Dtop^{NLO}(\kappa_{ugt})+(t+j)^{LO}$			& 2.413	&	3.283\\
$Dtop^{NLO}(\kappa_{cgt})+(t+j)^{LO}$			& 3.072	&	4.142\\
\hline
\end{tabular}
\end{center}
\caption{Efficiencies for the signal processes. \label{tab2}}
\end{table}

In table~\ref{tab2} we list the efficiencies for the signal processes after all cuts.
There is no significant difference between the LO and NLO samples in the ATLAS analysis. The only
notable difference arises in the ATLAS(m) analysis for the $Dtop^{NLO}+(t+j)^{LO}$ sample. As expected
the efficiency is better in ATLAS(m) than in ATLAS which is in accordance with the design of ATLAS(m). 
We now have to check if the rise in the number of signal events is enough to compensate for the increase
in the total background.

\begin{figure}[h!]
  \begin{center}
    \includegraphics[scale=0.50]{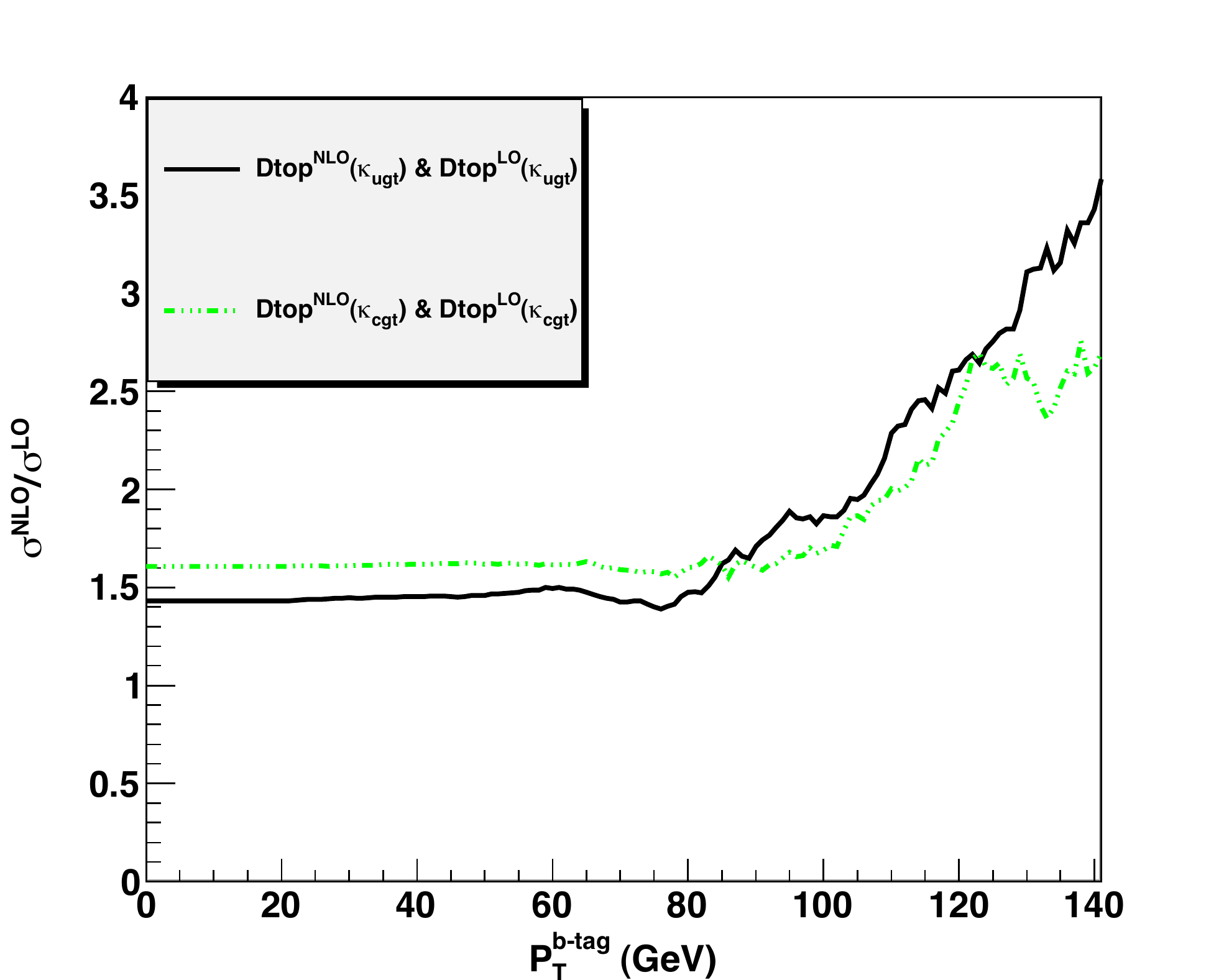}
    \caption{K-factor as function of the transverse momentum cut on the b-tagged jet.}
    \label{fig1}
  \end{center}
\end{figure}

As the LO and NLO results are quite similar, the NLO result seems to be well described by the LO sample 
with a constant K-factor.
In figure \ref{fig1} we plot the K-factor as a function of the transverse momentum cut of the b-tagged jet.
In this plot we have performed all cuts except the one on the b-jet in the ATLAS analysis. Then we have calculated
the ratio between the number of events coming from the  \textit{NLO} sample and the same number with the 
\textit{LO} sample for different values of the $p_T$ cut on the b-tagged jet. 
It is clear that the use of a constant K-factor is justified up to approximately a $p_T=60$ GeV cut. For large $p_T$
the recipe fails. However, the number of events decreases steeply with the b-jet $p_T$ cut for large
$p_T$ values and therefore their contribution to the total number of events becomes negligible. We have checked 
several other distributions always reaching the same conclusion - the regions where the use of a constant K-factor
would not be allowed, contribute with a small number of events to the analysis. Obviously, one should note that this is true for
this particular analysis and  not a general rule.

\section{Limits}

In this section we will present the bounds on the anomalous couplings for four different scenarios:
ATLAS analysis with the three samples $Dtop^{LO}$, $Dtop^{NLO}$ and $Dtop^{NLO}+(t+j)^{LO}$
and ATLAS(m) with $Dtop^{NLO}+(t+j)^{LO}$. In fact, because the ATLAS analysis with the 
$Dtop^{LO}$ and $Dtop^{NLO}$ leads to very similar results we will only show the results for
the NLO sample. Further, the ATLAS(m) analysis with only direct top events has
negligible signal events yields. 

As previously stated, the best current exclusion limit (assuming only one non-zero coupling at a time) 
was obtained in~\cite{Aad:2012gd} by ATLAS. With an energy of $\sqrt{s}=7\:TeV$ and a total integrated luminosity 
of $2.05\pm 0.08\: fb^{-1}$ the obtained limits at 95\% C.L. were
\begin{equation}
\kappa_{ugt}/\Lambda < 6.4\times 10^{-3}\:TeV^{-1} \quad \kappa_{cgt}/\Lambda < 14.5\times 10^{-3}\:TeV^{-1} \, \, . \label{eq:limit_atlas}
\end{equation}
As discussed, our goal is not to compete with this analysis but rather to understand if there is a way to improve it. 
According to our analysis there would be two possibilities to improve the bounds on the couplings. The first one
would be to just include the FCNC single top events in the signal, that is, to use the sample $Dtop^{NLO}+(t+j)^{LO}$.
The second would be to change the analysis by including an extra hard non-b jet (ATLAS(m)).
In order to obtain the 95\% CL limits for $\kappa_{ugt}/\Lambda$ and $\kappa_{cgt}/\Lambda$, we have used the code described 
in~\cite{ATLASthesis:2008106}.  

\begin{figure}[h!]
  \begin{center}
    \includegraphics[scale=0.42]{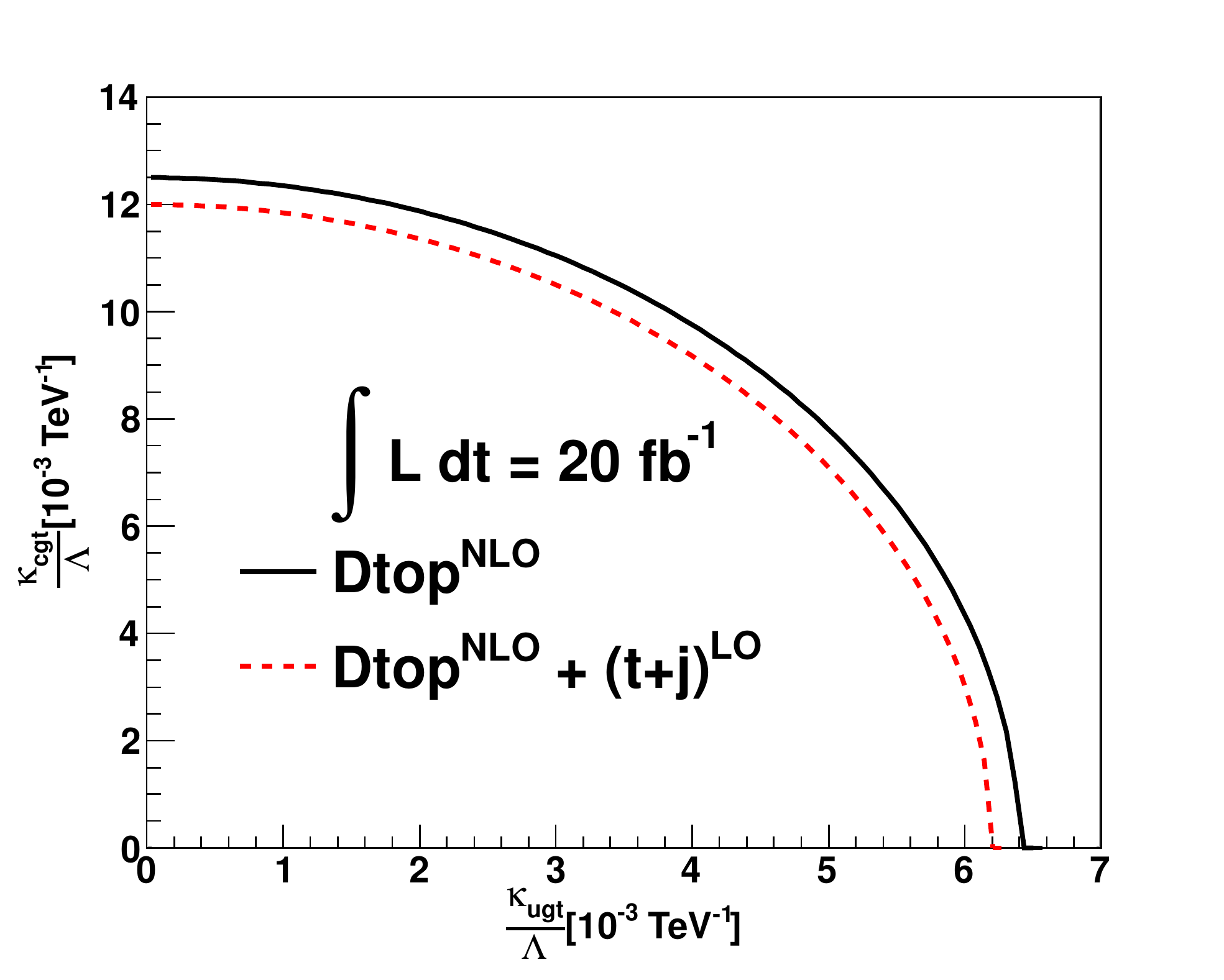} 
    \includegraphics[scale=0.42]{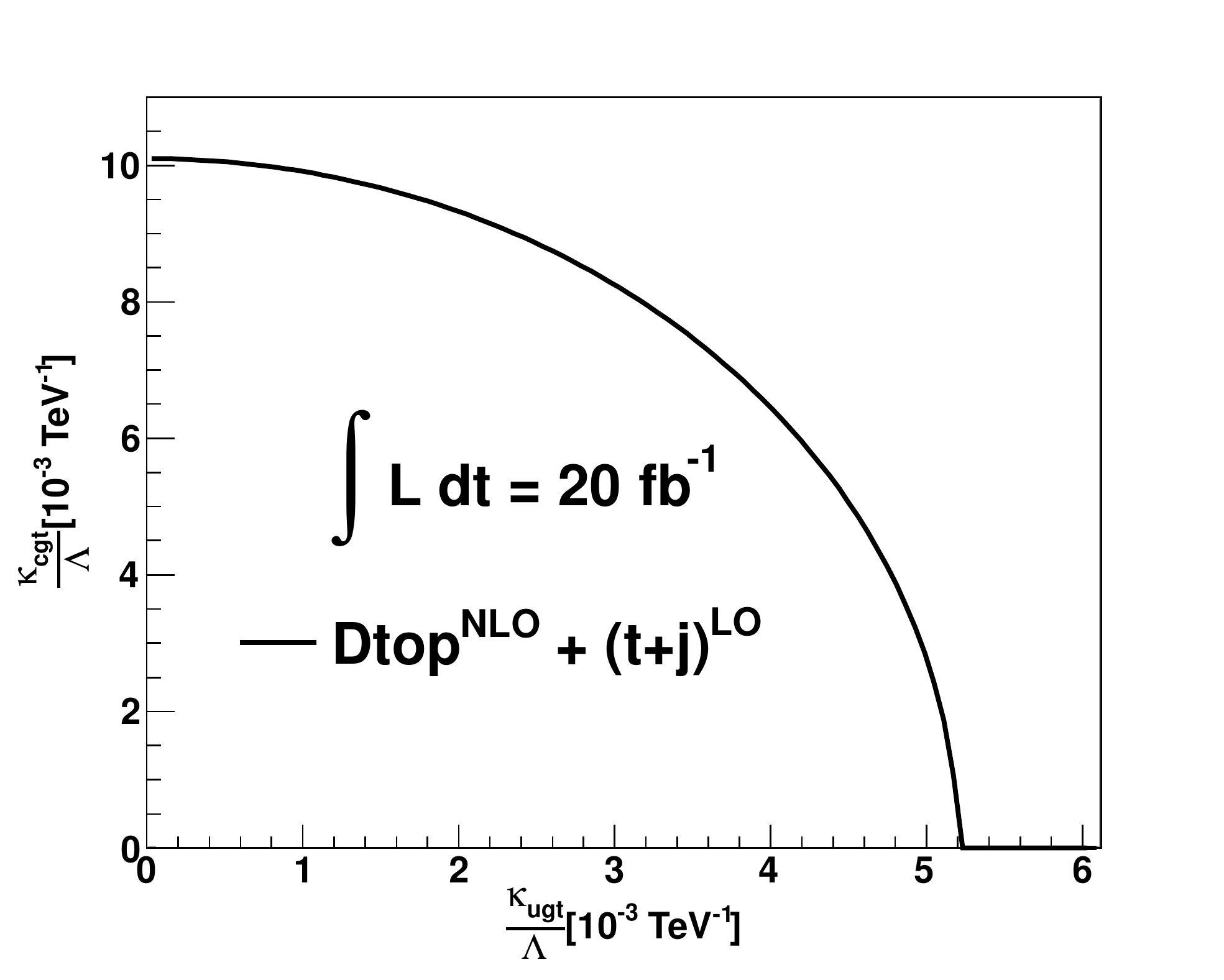}
    \caption{Left -Upper limit on the coupling constant $\kappa_{ugt}/\Lambda$ and $\kappa_{cgt}/\Lambda$ according the ATLAS analysis.
    Right - Upper limit on the coupling constant $\kappa_{ugt}/\Lambda$ and $\kappa_{cgt}/\Lambda$ according the ATLAS(m) analysis.}
    \label{fig3}
  \end{center}
\end{figure}

In fig.~\ref{fig3} we present the results for $\sqrt{s}=8\:TeV$ and a total integrated luminosity of $20\: fb^{-1}$.
In the left panel we show the 95\% C.L. upper limit on the coupling constant $\kappa_{ugt}/\Lambda$ and 
$\kappa_{cgt}/\Lambda$ according to the ATLAS analysis for the $Dtop^{NLO}$ sample (black line) and for
the $Dtop^{NLO}+(t+j)^{LO}$ sample (slashed red line). The LO result with the K-factor is almost
on top of the the NLO one and therefore it is not shown in the figure. Hence, for this particular analysis,
no significant difference is seen when using the NLO generator.
In the right panel we show the ATLAS(m) analysis
with the $Dtop^{NLO}+(t+j)^{LO}$ sample. The numeric results for the limits with each coupling taken non-zero at a time
are shown in table~\ref{tab3} (they correspond to the intersection of the exclusion curve with the x and y axes).

\begin{table}[h!]
\begin{center}
\begin{tabular}{l|c|c}
\hline
  & ATLAS & ATLAS(m) \\
\hline
Process   & $\kappa/\Lambda\: [10^{-3}]\: TeV^{-1}$ &  $\kappa/\Lambda\:
[10^{-3}]\: TeV$ \\
\hline
$Dtop^{NLO}(\kappa_{ugt})$                                        &  6.4     &       --     \\
$Dtop^{NLO}(\kappa_{cgt})$                                        & 12.5     &       --     \\
$Dtop^{NLO}(\kappa_{ugt})+(t+j)^{LO}$               &  6.2     &      5.2     \\
$Dtop^{NLO}(\kappa_{cgt})+(t+j)^{LO}$               & 12.0     &     10.1     \\
\hline
\end{tabular}
\end{center}
\caption{Limits of the $\kappa_{qgt}/\Lambda.$ \label{tab3}}
\end{table} 

The results presented in fig.~\ref{fig3} and in table~\ref{tab3}  allow us to conclude that a slight improvement
in the bound can be achieved by including the full set of events $Dtop^{NLO}+(t+j)^{LO}$  in the ATLAS analysis.
The same is true for the ATLAS(m) analysis when compared with ATLAS, even when the $Dtop^{NLO}+(t+j)^{LO}$ 
is considered. However, we should look at this results as an indication that a full detector analysis is worth doing.
First because the improvement is only of the order a few \% and second because since we did not consider
the multi-jet background, we could be overestimating the sensitivity in ATLAS(m).


%

\section{Conclusions}
We have studied top production at the LHC via FCNC interactions. We have performed two different
analysis using the MEtop generator which allows for the production of NLO direct top events and LO FCNC $pp \to t j$ events.
In the first one we have followed ATLAS in~\cite{Aad:2012gd} but using two different samples, one with
only direct top NLO events and the other one with direct top NLO plus the LO $t+j$ FCNC events. In the second
analysis we allow for an extra hard non-b jet in the final state increasing both the signal and the background
yields.


We have concluded that the inclusion of the FCNC single top events has increased the sensitivity. 
Even if the limit is better by only a few percent this should be implemented in the experimental analysis 
because this is a real contribution to the process and should not be neglected. Furthermore, its
inclusion is straightforward with MEtop. In the second analysis the limit on the couplings is significantly better. 
In this case however we 
should look at the result as an indication that an experimental analysis is worth performing.


\begin{thebibliography}{99}

\bibitem{AguilarSaavedra:2004wm} 
  J.~A.~Aguilar-Saavedra,
  Acta Phys.\ Polon.\ B {\bf 35}, 2695 (2004)
  [hep-ph/0409342].
  
\bibitem{Aaltonen:2008qr} 
  T.~Aaltonen {\it et al.}  [CDF Collaboration],
  Phys.\ Rev.\ Lett.\  {\bf 102}, 151801 (2009)
  [arXiv:0812.3400 [hep-ex]];
%
  V.~M.~Abazov {\it et al.}  [D0 Collaboration],
  Phys.\ Lett.\ B {\bf 693}, 81 (2010)
  [arXiv:1006.3575 [hep-ex]].

\bibitem{Aad:2012gd} 
  G.~Aad {\it et al.}  [ATLAS Collaboration],
  Phys.\ Lett.\ B {\bf 712}, 351 (2012)
  [arXiv:1203.0529 [hep-ex]].

\bibitem{Saavedra:2013} 
  J. A. Aguilar-Saavedra,
  ``Protos: Program for Top Simulation (User Manual, version 2.1),''
	on http:\/\/jaguilar.web.cern.ch\/jaguilar\/protos\/

\bibitem{Liu:2005dp} 
  J.~J.~Liu, C.~S.~Li, L.~L.~Yang and L.~G.~Jin,
  Phys.\ Rev.\ D {\bf 72}, 074018 (2005)
  [hep-ph/0508016].

\bibitem{Gao:2011fx} 
  J.~Gao, C.~S.~Li, L.~L.~Yang and H.~Zhang,
  Phys.\ Rev.\ Lett.\  {\bf 107}, 092002 (2011)
  [arXiv:1104.4945 [hep-ph]].

\bibitem{Coimbra:2012ys} 
  R.~Coimbra, A.~Onofre, R.~Santos and M.~Won,
  Eur.\ Phys.\ J.\ C {\bf 72}, 2222 (2012)
  [arXiv:1207.7026 [hep-ph]];
  R.~Coimbra, A.~Onofre, R.~Santos and M.~Won,
  J.\ Phys.\ Conf.\ Ser.\  {\bf 447}, 012031 (2013)
  [arXiv:1303.5485 [hep-ph]].

\bibitem{alot}
  P.~M.~Ferreira, O.~Oliveira and R.~Santos,
  Phys.\ Rev.\  D {\bf 73} (2006) 034011
  [arXiv:hep-ph/0510087];
%
  P.~M.~Ferreira and R.~Santos,
  Phys.\ Rev.\  D {\bf 73} (2006) 054025
  [arXiv:hep-ph/0601078];
%
%
  R.~A.~Coimbra, P.~M.~Ferreira, R.~B.~Guedes, O.~Oliveira, A.~Onofre, R.~Santos and M.~Won,
  Phys.\ Rev.\  D {\bf 79} (2009) 014006
  [arXiv:0811.1743 [hep-ph]];
%
  J.~A.~Aguilar-Saavedra,
  Nucl.\ Phys.\  {\bf B812}, 181-204 (2009).
  [arXiv:0811.3842 [hep-ph]];
%
  B.~Grzadkowski, M.~Iskrzynski, M.~Misiak, J.~Rosiek,
  JHEP {\bf 1010 } (2010)  085.
  [arXiv:1008.4884 [hep-ph]];
%
  J.~A.~Aguilar-Saavedra,
  Nucl.\ Phys.\  B {\bf 843} (2011) 638
  [Erratum-ibid.\  B {\bf 851} (2011) 443]
  [arXiv:1008.3562 [hep-ph]].

\bibitem{Pumplin:2002vw} 
  J.~Pumplin, D.~R.~Stump, J.~Huston, H.~L.~Lai, P.~M.~Nadolsky and W.~K.~Tung,
  JHEP {\bf 0207}, 012 (2002)
  [hep-ph/0201195].

\bibitem{Alioli:2010xd}
  S.~Alioli, P.~Nason, C.~Oleari and E.~Re,
  JHEP {\bf 1006} (2010) 043;
%
S.~Alioli, P.~Nason, C.~Oleari and E.~Re,
 JHEP {\bf 0909} (2009) 111
 [Erratum-ibid.\ {\bf 1002} (2010) 011];
%
%
E.~Re,
   Eur.\ Phys.\ J.\ C {\bf 71} (2011) 1547;
%
%
 S.~Frixione, P.~Nason and G.~Ridolfi,
 JHEP {\bf 0709} (2007) 126;
%
%
 P.~Nason,
   JHEP {\bf 0411} (2004) 040;
%
 S.~Frixione, P.~Nason and C.~Oleari,
   JHEP {\bf 0711} (2007) 070.

\bibitem{Mangano:2002ea}
  M.~L.~Mangano, M.~Moretti, F.~Piccinini, R.~Pittau and A.~D.~Polosa,
  JHEP {\bf 0307} (2003) 001.

\bibitem{mlm} M. Mangano, \textit{Merging multijet matrix elements
and shower evolution in hadronic collisions,} http://cern.ch/\%7Emlm/talks/lund-alpgen.pdf (2004).

\bibitem{Campbell:1999ah}
  J.~M.~Campbell and R.~K.~Ellis,
  Phys.\ Rev.\  D {\bf 60} (1999) 113006.

\bibitem{site}
Code available from http://mcfm.fnal.gov/.


\bibitem{Sjostrand:2006za}
  T.~Sjostrand, S.~Mrenna and P.~Z.~Skands,
  JHEP {\bf 0605} (2006) 026.

\bibitem{Skands:2009zm} 
  P.~Z.~Skands,
  arXiv:0905.3418 [hep-ph].

\bibitem{ATLASCollaboration:2010002} 
  {\it ``ATLAS Monte Carlo tunes for MC09,''}
  ATL-PHYS-PUB-2010-002.

\bibitem{Ovyn:2009tx}
  S.~Ovyn, X.~Rouby and V.~Lemaitre,
  arXiv:0903.2225 [hep-ph].


\bibitem{ATLASthesis:2008106}
  F.~M.~A.~Veloso, J.~Carvalho and A.~Onofre, 
  {\it ``Study of ATLAS sensitivity to FCNC top quark decays,''}
  CERN-THESIS-2008-106.

\end{thebibliography}
\end{document}